\documentclass[%
reprint,
superscriptaddress,
amsmath,amssymb,
aps,
]{revtex4-2}
\usepackage{graphicx}
\usepackage{dcolumn}
\usepackage{bm}
\usepackage{xcolor}
\usepackage{siunitx}
\usepackage{array}
\usepackage{booktabs}
\usepackage{multirow}
\usepackage{dcolumn}

\begin{document}

\title{Quantum nonreciprocity from qubits coupled by Dzyaloshinskii–Moriya interaction}
\author{Zhenghao Zhang}
\email{zzh312@tamu.edu}
\author{Qingtian Miao}
\email{qm8@tamu.edu}
\thanks{Z.Z. and Q.M. contributed equally to this work.}
  \affiliation{Department of Physics and Astronomy, Texas A\&M University, Texas 77843, USA}
\author{G. S. Agarwal}
\email{Girish.Agarwal@ag.tamu.edu}
 \affiliation{Institute for Quantum Science and Engineering, Department of Biological and Agricultural Engineering, Department of Physics and Astronomy, Texas A\&M University, College Station, Texas 77843, USA}

\date{\today}

\begin{abstract}
We present a theoretical study of quantum nonreciprocity induced via a Dzyaloshinskii–Moriya interaction (DMI) in an, otherwise achiral, waveguide quantum electrodynamics. Using the full quantum master equation and input–output formalism for two two-level systems coupled to a one-dimensional waveguide and driven by a coherent field, we show that an engineered DMI enables strong nonreciprocity in an otherwise reciprocal system, with tunable behavior governed by driving strength, detuning, and phase of the DMI. We not only demonstrate nonreciprocal transmission but also demonstrate nonreciprocal quantum entanglement and photon bunching. The system can end up in a pure state as certain decohering channels do not participate. The pure state leads to power-independent perfect transparency. Conditions are derived and depend on the propagation phase, the relative detuning of the two qubits, and the exchange interaction. At these pure-state points, the steady-state entanglement is reciprocal and admits a closed-form expression; away from them, phase control generates strong entanglement nonreciprocity. The DMI also reshapes photon statistics, redistributing two-photon correlations and shifting superbunching from transmission (no DMI) to reflection at finite DMI. These results establish DMIs as a versatile resource for engineering nonreciprocity, transparency, entanglement, and photon correlations in waveguide QED, enabling isolators, routers, and superbunching light sources without requiring chiral waveguides.
\end{abstract}

\maketitle

\section{Introduction}

Nonreciprocity—unequal forward/backward response between the same modes—requires breaking Lorentz reciprocity; in linear, time-invariant, energy-conserving systems, this is tantamount to breaking time-reversal symmetry \cite{carminati2000reciprocity,caloz2018electromagnetic,zhao2019connection,asadchy2020tutorial}. Techniques that break this symmetry are central to signal routing and sensing, including magneto-optic bias \cite{bi2011chip,shoji2014magneto} and, without magnets, spatiotemporal modulation \cite{lira2012electrically, fang2012photonic, sounas2017non}, synthetic gauge fields \cite{fang2012realizing,fang2013controlling,biehs2023enhancement,estep2014magnetic,rosen2024synthetic}, optomechanical interactions \cite{metelmann2015nonreciprocal,malz2018quantum,xu2019nonreciprocal,mirhosseini2020superconducting} or engineered nonlinearity (nonlinear PT-symmetric media \cite{peng2014parity,liu2014regularization,chang2014parity}, stimulated Brillouin scattering \cite{dong2015brillouin,kim2015non}, Kerr nonlinearity \cite{cotrufo2021nonlinearity1,cotrufo2021nonlinearity2,miao2024kerr}, etc.). There is growing interest in pushing nonreciprocity into the quantum regime. As resources for quantum networking and state transfer, chiral light–matter interfaces provide directional coupling \cite{stannigel2012driven,pichler2015quantum,lodahl2017chiral,owens2022chiral}, nonlinearities enable one-way quantum processors \cite{rosario2018nonreciprocity,xia2018cavity,tang2022quantum,chen2023nonreciprocal,chen2024nonreciprocal}, and topological channels \cite{yao2013topologically,lemonde2019quantum,ozawa2019topological,de2023chiral} yield disorder-robust routing of quantum states. As a metrological primitive, non-Hermitian/nonreciprocal architectures offer routes to enhanced sensing \cite{lau2018fundamental,chen2019sensitivity,mcdonald2020exponentially}. Quantum nonreciprocity can be realized across superconducting parametric circuits \cite{kamal2011noiseless,abdo2014josephson,roushan2017chiral,lecocq2017nonreciprocal,wang2019synthesis,liu2020synthesizing,joshi2023resonance,almanakly2025deterministic}, optomechanical platforms \cite{jiao2020nonreciprocal,jiao2022nonreciprocal,chen2023nonreciprocal}, and spinning-resonator optics \cite{maayani2018flying,huang2018nonreciprocal,chen2024nonreciprocal}. It is worth noting that recent theory shows that nonreciprocity qualitatively reshapes many-body interactions and critical behavior \cite{chiacchio2023nonreciprocal, zhu2024nonreciprocal,brighi2024nonreciprocal,begg2024quantum}.

Unlike free-space systems, chiral waveguide QED provides structured environments that enable directional photon–emitter coupling \cite{pucher2022atomic,liedl2024observation} and dissipation-driven formation of entangled states among emitters \cite{stannigel2012driven,ramos2014quantum,pichler2015quantum,gonzalez2015chiral,mirza2016multiqubit,mirza2016two}. Building on our recent analysis of two detuned two-level systems (qubits) coupled to a chiral waveguide—where we identified the parameter regimes for transparency and quantum nonreciprocity in transmission and fluctuations \cite{miao2025transparency}—we now ask whether comparable behavior can be realized without chiral couplers. We show that a nonchiral waveguide suffices if the qubits are linked by a complex inter-qubit exchange $\mathcal J$. Formally, a complex exchange can be decomposed into a symmetric (real) exchange and an antisymmetric contribution, i.e., the Dzyaloshinskii–Moriya interaction \cite{dzyaloshinskii1957thermodynamic,moriya1960anisotropic}, such as in 3D noncentrosymmetric metals \cite{wang2017rkky}. A concrete realization is a magnetic bilayer separated by a nonmagnetic spacer, where the spacer mediates a complex coupling whose antisymmetric contribution appears only when the spacer breaks inversion symmetry \cite{Zakeri2010Asymmetric,moon2013spin,zou2024dissipative}. In our work, we use “DMI” to denote the corresponding effective antisymmetric exchange term between two pseudo-spin-1/2 qubits. Synthetic realizations also exist in superconducting parametric circuits, where synthetic gauge fields and Floquet control implement DM-like couplings, chiral excitation flow, and multi-spin chirality in qubit arrays \cite{roushan2017chiral,wang2019synthesis,liu2020synthesizing}. In particular, Wang et al. synthesized antisymmetric exchange by phase-patterned periodic modulation of qubit frequencies in a five-qubit superconducting circuit, realizing a three-qubit chiral gate and preparing Greenberger–Horne–Zeilinger (GHZ) states involving up to five qubits \cite{wang2019synthesis}. In that platform, each qubit’s transition frequency is individually tunable via its Z bias line, and the engineered exchange is controlled by the amplitudes and relative phases of the applied frequency modulations.

Our paper is organized as follows. In Sec. \ref{modelSec} we formulate a two-qubit waveguide QED model with a complex exchange $\mathcal{J}=J e^{i\theta}$, together with the full master equation and input–output relations. The antisymmetric DMI part in the exchange contributes a phase-biased inter-qubit coupling that skews left–right response despite a nonchiral photonic environment. In Sec. \ref{NonreSec} we investigate nonreciprocity in transport—coherent and incoherent components of transmission—arising from the antisymmetric exchange. In Sec. \ref{transSec} we analyze pure steady states: for phase-matched separations $\phi=n\pi$ we derive existence conditions under symmetric and antisymmetric detunings, and show that these states yield reciprocal transparency. Section \ref{entSec} maps nonreciprocal steady-state entanglement, using concurrence to reveal how the magnitude and phase of the DMI make forward/backward contrasts beyond the pure-state points. Section \ref{SuperbunchSec} treats photon statistics and demonstrates that the DMI redistributes two-photon correlations, shifting superbunching from transmission (no DMI) to reflection at finite DMI. We conclude in Sec. \ref{conclusion}.

\section{MODEL and basic equations} \label{modelSec}
\begin{figure}[htbp]
    \includegraphics[width=0.48\textwidth]{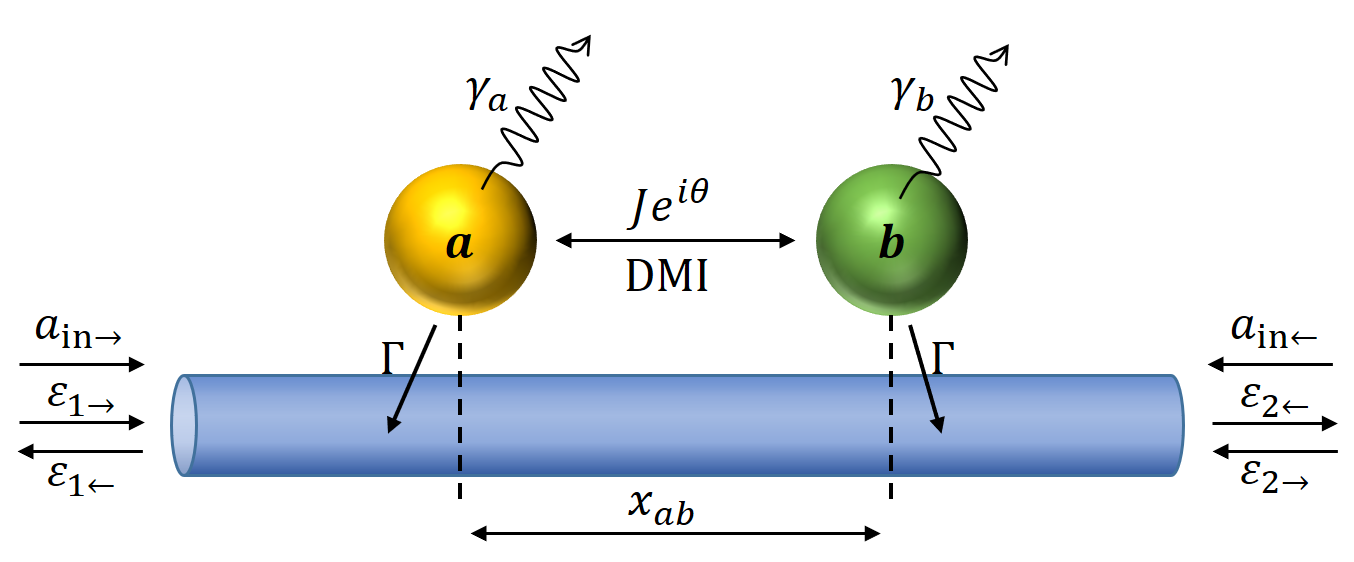}
    \caption{Two qubits $a,b$ separated by $x_{ab}$ along a bidirectional 1D waveguide. Each qubit couples to the waveguide at rate $\Gamma$ and to other loss channels at rate $\gamma$. The inter-qubit exchange is complex, $\mathcal{J}=J e^{i\theta}$, whose DMI part $J\sin\theta$ produces an $i(S_a^+S_b^--S_b^+S_a^-)$ coupling. A coherent drive at frequency $\omega_d$ is injected from the left (port 1, right-going) or from the right (port 2, left-going) with amplitudes $\varepsilon_{1\rightarrow}$ and $\varepsilon_{2\leftarrow}$. Propagation between the qubits adds a phase $\phi = \omega_d x_{ab}/v_p$, with $v_p$ denoting the phase velocity of the drive. Quantum vacuum input fields are $(a_{\rm in\rightarrow},a_{\rm in\leftarrow})$.}
    \label{model}
\end{figure}
As shown in Fig.~\ref{model}, two qubits $a,b$ are separated by $x_{ab}$ along a bidirectional nonchiral 1D waveguide and driven at frequency $\omega_d$. The left- and right-propagating fields are coupled symmetrically to each qubit. We take $\Gamma$ to denote the decay rate into the waveguide, and $\gamma_a$, $\gamma_b$ as additional loss to non-waveguide channels. We work in a frame rotating at $\omega_d$ ($\hbar=1$), define detunings $\Delta_j=\omega_j-\omega_d$ ($j\in\{a,b\}$) and a propagation phase $\phi=\omega_d x_{ab}/v_p$, and use spin-$\tfrac12$ operators $S_j^z=\tfrac12(|e\rangle\!\langle e|-|g\rangle\!\langle g|)$ and $S_j^\pm$.

The qubits are linked by a complex inter-qubit exchange $\mathcal{J}=J e^{i\theta}$ and driven from the left (port 1, right-going) or from the right (port 2, left-going) with coherent amplitudes $\varepsilon_{1\rightarrow}$ and $\varepsilon_{2\leftarrow}$. Without loss of generality we take $J\ge 0$ and restrict the DMI phase to $\theta\in[0,2\pi)$. The Hamiltonian is
\begin{align}
    H &= \Delta_a S^z_a + \Delta_b S^z_b + (J e^{i\theta} S_a^+ S_b^- + J e^{-i\theta} S_a^- S_b^+) \notag\\
    &+ (i k \varepsilon_{1\rightarrow} S^+_a - i k^* \varepsilon_{1\rightarrow}^* S^-_a) \notag\\
    &+ (i k e^{i\phi} \varepsilon_{1\rightarrow} S^+_b - i k^* e^{-i\phi} \varepsilon_{1\rightarrow}^* S^-_b) \notag\\
    &+ (i k e^{i\phi} \varepsilon_{2\leftarrow} S^+_a - i k^* e^{-i\phi} \varepsilon_{2\leftarrow}^* S^-_a) \notag\\
    &+ (i k \varepsilon_{2\leftarrow} S^+_b - i k^* \varepsilon_{2\leftarrow}^* S^-_b).
    \label{Hamiltonian}
\end{align} 
where $k$ denotes the qubit–waveguide coupling strength. Note that when $\theta\neq n\pi$, the complex exchange term breaks time-reversal symmetry as discussed in Appendix \ref{timerev}.

Within the standard Born–Markov approximations, a master equation for the two-qubit density operator can be written as
\begin{align}
    \dot\rho &= -i[H,\rho]-\sum_{i=a,b}(\Gamma+\gamma_i)(S^+_i S^-_i \rho - 2S^-_i \rho S^+_i  \notag\\ 
    &+\rho S^+_i S^-_i) 
    - \Gamma[e^{i\phi} S^+_a S^-_b \rho - (e^{i\phi}+e^{-i\phi}) S^-_b \rho S^+_a \notag\\
    &+ e^{-i\phi} \rho S^+_a S^-_b + H.c.],
    \label{master}
\end{align}
with $\Gamma=|k|^2$, consistent with the independent derivation in Ref.~\cite{pichler2015quantum,mukhopadhyay2022anti,miao2025transparency}. The corresponding port quantum fields obey the input–output relations
\begin{align}
    \varepsilon_{1\leftarrow}  &= e^{i\phi} \varepsilon_{2\leftarrow}  -k^* (S_a^- + e^{i\phi}S_b^-), \notag\\
    \varepsilon_{2\rightarrow} &= e^{i\phi} \varepsilon_{1\rightarrow} -k^*(e^{i\phi}S_a^- + S_b^-).
    \label{inoutput}
\end{align}
In writing Eqs. (\ref{inoutput}), we drop the vacuum input fields as these do not contribute to normally ordered expectation values. All numerical results are obtained by solving the master equation, Eq.~(\ref{master}). The resulting steady-state density matrix elements are then used to evaluate expectation values of the transmitted and reflected fields via the input–output relations, Eq.~(\ref{inoutput}). Equations~(\ref{master}) and (\ref{inoutput}) therefore constitute the basic framework of this work. Importantly, within this framework, the forward-transmission response is mapped to the backward-transmission response under $\theta\to-\theta$ when $\Delta_a=\Delta_b$; details are provided in Appendix~\ref{qubitex}. 

Using Eq.~\eqref{master} one obtains equations of motion for the atomic dipole amplitudes
\begin{align}
\langle \dot S_a^- \rangle &= -(\Gamma+\gamma_a+i\Delta_a)\langle S_a^- \rangle
+ \Bigl[i J e^{i\theta} + \Gamma e^{i\phi}\Bigr]\langle 2 S_a^z S_b^- \rangle \notag\\
&\quad - k\Bigl[\varepsilon_{1\rightarrow}+e^{i\phi}\varepsilon_{2\leftarrow}\Bigr]\langle 2 S_a^z\rangle, \notag\\
\langle \dot S_b^- \rangle &= -(\Gamma+\gamma_b+i\Delta_b)\langle S_b^- \rangle
+ \Bigl[i J e^{-i\theta} + \Gamma e^{i\phi}\Bigr]\langle 2 S_b^z S_a^- \rangle \notag\\
&\quad - k\Bigl[\varepsilon_{2\leftarrow}+e^{i\phi}\varepsilon_{1\rightarrow}\Bigr]\langle 2 S_b^z\rangle.
\label{eq:cmt_dyn}
\end{align}
\begin{figure*}[t]
  \includegraphics[width=0.96\textwidth]{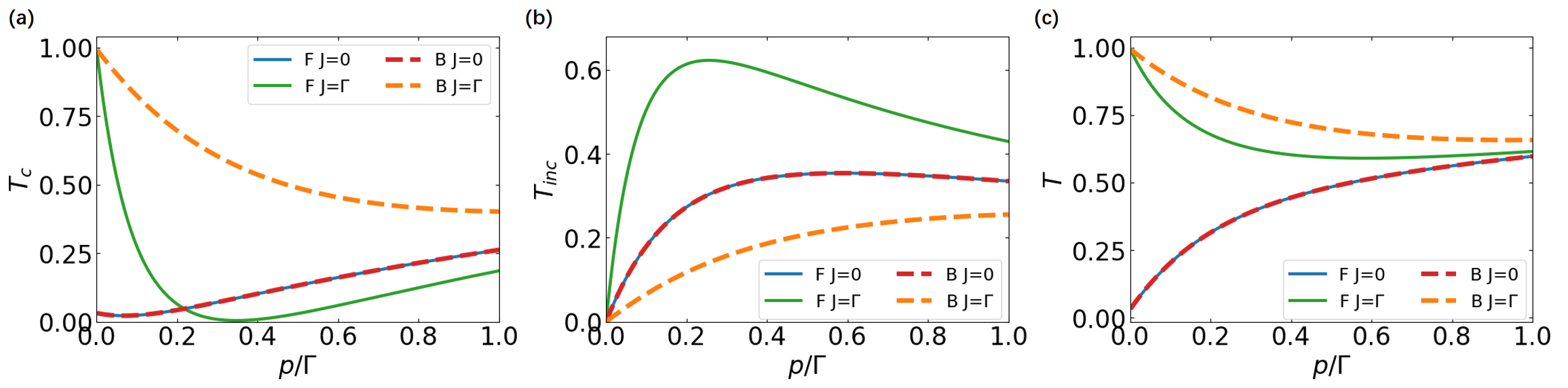}
  \caption{Forward (F, solid) and backward (B, dashed) transmission under single-sided driving (the counterpropagating input is zero). Panels: (a) coherent component \(T_{\mathrm c}\), (b) incoherent component \(T_{\mathrm inc}\), and (c) total transmission \(T=T_{\mathrm c}+T_{\mathrm inc}\) versus drive power \(p/\Gamma\). Curves compare a reciprocal reference (\(\mathcal{J}=0\)) with a complex exchange (\(\mathcal{J}=\Gamma e^{18i\pi/25}\)). Parameters: \(\Delta_a=\Delta_b=0.5\,\Gamma\), \(\phi=9\pi/25\). These phases are chosen to produce pronounced nonreciprocity. All curves show steady-state quantities obtained by numerically solving the master equation [Eq. (\ref{master})]. Unless stated otherwise, since our normalized quantities are divided by the input power \(p\), we start the $p$-axis at \(p=10^{-3}\Gamma\).}
  \label{nonre}
\end{figure*}

In the weak-drive (linear-response) regime we can take $\langle S^z_j\rangle\simeq-1/2$ and apply a mean-field factorization of correlators, which reproduces the semiclassical equations in \cite{miao2024kerr}. Defining the inter-qubit couplings
\begin{equation}
A_{b\rightarrow a}= i J e^{i\theta} + \Gamma e^{i\phi},\qquad A_{a\rightarrow b}= i J e^{-i\theta} + \Gamma e^{i\phi},
\end{equation}
reveals the interaction imbalance \(A_{b\rightarrow a}-A_{a\rightarrow b}=2J\sin\theta\), so a finite DMI phase $\theta\neq0,\pi$ biases the $a\to b$ and $b\to a$ pathways and breaks left–right symmetry even though the waveguide coupling is nonchiral. A particularly instructive tuning cancels one pathway. For $J=\Gamma$, $\theta=\phi+\tfrac{\pi}{2}\ (\bmod 2\pi)$, $A_{b\rightarrow a}=0$ while $A_{a\rightarrow b}\neq0$, realizing a one-way coupling $a\to b$. Conversely, cancelling the opposite direction requires $J=\Gamma$, $\theta=-\phi+\tfrac{\pi}{2}\ (\bmod 2\pi)$ which gives $A_{a\rightarrow b}=0$ while $A_{b\rightarrow a}\neq0$. A comprehensive semiclassical discussion of how the DMI phase $\theta$ and propagation phase $\phi$ cooperate to generate nonreciprocity is given in Ref.~\cite{miao2024kerr,zou2024dissipative}. In the fully quantum regime the same bias seeds nonreciprocal scattering—affecting both transmission, entanglement and photon correlations—which we analyze below. For simplicity we set $\gamma_a=\gamma_b=0$ and $k$ real.

\section{Quantum nonreciprocal transmission induced by DMI}
\label{NonreSec}
\begin{figure*}[htbp]
  \includegraphics[width=0.5\textwidth]{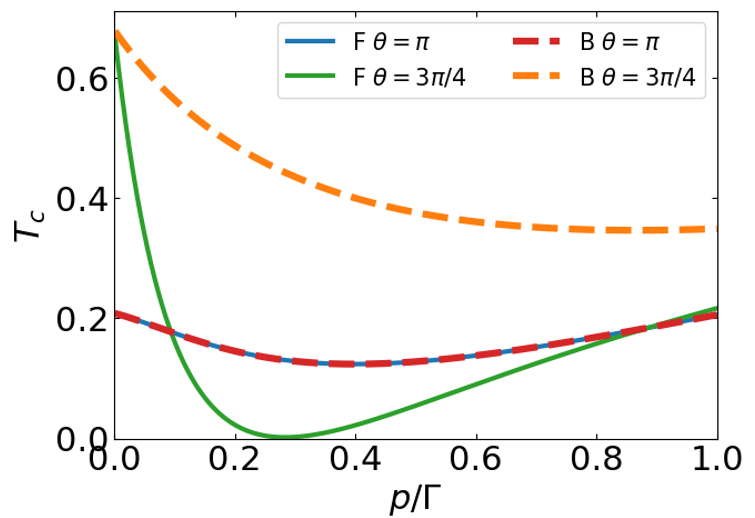}
  \caption{Forward (F, solid) and backward (B, dashed) coherent transmission versus drive power \(p/\Gamma\) under single-sided driving (the counterpropagating input is zero). Curves compare with or without DMI  (\(\theta=\pi,3\pi/4\)). Parameters: \(\Delta_a=\Delta_b=0.5\,\Gamma\), \(\phi=\pi/4\), \(|J|=\Gamma\).}
  \label{nonre3}
\end{figure*}
Nonreciprocity in a nonchiral waveguide has been observed by exploiting intrinsic nonlinearity of two-level systems together with asymmetric detuning, which breaks the structural symmetry and yields nonreciprocity (e.g., Hamann \textit{et al.}~\cite{rosario2018nonreciprocity}). Here we use a different mechanism: an engineered DMI between the two qubits. The DMI introduces a phase bias in the inter-qubit interactions, so that the forward and backward scattering pathways are intrinsically unbalanced even with symmetric detunings. By contrast, Ref.~\cite{rosario2018nonreciprocity} operates with no exchange ($\mathcal J=0$) and relies on a propagation phase $\phi$ tuned near but not exactly at $\pi$.

We probe nonreciprocity under single-port driving and analyze the long-time steady state ($\dot\rho=0$). For forward drive we set $\varepsilon_{1\rightarrow}=\alpha$ (real, $\alpha\neq0$) and $\varepsilon_{2\leftarrow}=0$, and define input power $p\equiv \alpha^{2}$. Backward drive exchanges the ports $(\varepsilon_{1\rightarrow},\varepsilon_{2\leftarrow})=(0,\alpha)$. Normalizing by $p$, we define the transmitted and reflected intensities
\begin{equation}
T=\frac{\langle \varepsilon_{2\rightarrow}^\dagger \varepsilon_{2\rightarrow}\rangle}{p},\qquad
R=\frac{\langle \varepsilon_{1\leftarrow}^\dagger \varepsilon_{1\leftarrow}\rangle}{p},
\end{equation}
and separate them into coherent and incoherent parts,
\begin{equation}
\begin{aligned}
T_{\rm c}=\frac{|\langle \varepsilon_{2\rightarrow}\rangle|^{2}}{p}&,\quad
R_{\rm c}=\frac{|\langle \varepsilon_{1\leftarrow}\rangle|^{2}}{p},\\
T_{\rm inc}=T-T_{\rm c}&,\quad
R_{\rm inc}=R-R_{\rm c}.
\end{aligned}
\end{equation}
In the figures we focus on transmission and its decomposition, $T=T_{\rm c}+T_{\rm inc}$.

Fig.~\ref{nonre} shows transmission under single-port drive for a reciprocal reference ($\mathcal J=0$) and for a complex exchange ($\mathcal{J}=\Gamma e^{18i\pi/25}$) at symmetric detuning ($\Delta_a=\Delta_b$). 
For $\mathcal J=0$, the forward (F) and backward (B) traces coincide in all panels, as required by structural symmetry. Introducing the DMI yields a clear direction-dependent splitting between the F and B curves:

(a) \emph{Coherent transmission $T_{\rm c}$.} 
In the weak-drive limit ($p\!\to\!0$) the incoherent part vanishes due to negligible quantum nonlinearity. 
Without exchange the coherent part in transmission is near zero for the chosen parameters, whereas with exchange it can approach unity ($T_{\rm c}\simeq1$). Further details are provided in Appendix~\ref{weaklimit}. As $p$ increases, the forward curve $T_{\rm c}^{\rm F}$ (green, solid) drops sharply and nearly vanishes ($T_{\rm c}\simeq0.004$) near $p/\Gamma\simeq0.35$, while the backward curve $T_{\rm c}^{\rm B}$ (orange, dashed) stays larger than the reciprocal reference across the full range.

(b) \emph{Incoherent transmission $T_{\rm inc}$.} 
With DMI, $T_{\rm inc}^{\rm F}$ (green, solid) is enhanced relative to $\mathcal J=0$, whereas $T_{\rm inc}^{\rm B}$ (orange, dashed) is suppressed. This forward–backward separation persists over the entire power window plotted.

(c) \emph{Total transmission $T=T_{\rm c}+T_{\rm inc}$.}
At weak drive, the DMI curves deviate markedly from the reciprocal reference, while remaining close to each other. As $p$ increases, the forward and backward totals both approach the $\mathcal J=0$ trace.

A key control knob is the phase pair ($\phi$, $\theta$). Choosing $\phi=9\pi/25$ and $\theta=18\pi/25$ produces a pronounced directional contrast: in one direction the coherent component is suppressed while the incoherent component is enhanced, and in the opposite direction the coherent part is enhanced with the incoherent part suppressed. Thus ($\phi$, $\theta$) program whether transmission is carried predominantly by elastic or inelastic scattering in each direction, enabling tunable quantum nonreciprocity in an otherwise reciprocal waveguide.

To isolate the role of DMI in governing nonreciprocity, we perform a benchmark at fixed $J\neq 0$ by comparing
(i) $\theta=0$ or $\pi$, for which the coupling is purely real and the DMI component vanishes
($J\sin\theta=0$), with (ii) $\sin\theta\neq 0$, for which a finite DMI is present. The results are shown in Fig.~\ref{nonre3}. For purely real exchange (e.g., $\theta=\pi$; red and blue curves),
the forward and backward responses coincide, i.e., the transmission is reciprocal.
By contrast, for a finite DMI (e.g., $\theta=3\pi/4$; red and blue curves), the forward and backward
transmissions separate markedly once the system enters the nonlinear (finite-power) regime, producing
pronounced nonreciprocity. This benchmark makes explicit that the directional effects reported in the main
text are controlled by the DMI component $J\sin\theta$, whereas the purely real exchange $J\cos\theta$
does not by itself generate directional asymmetry in our setup.

\begin{figure*}[tbp]
    \includegraphics[width=0.96\textwidth]{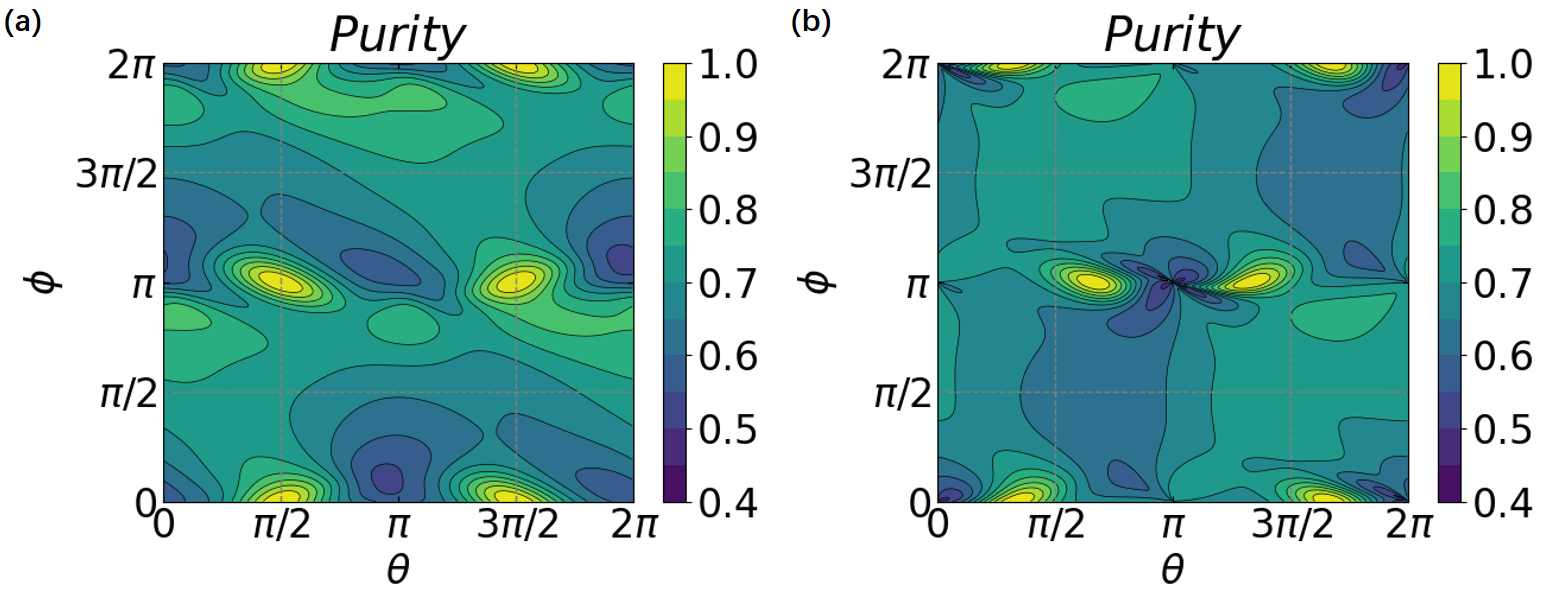}
    \caption{The Purity $\mathrm{Tr}[\rho^2]$ as a function of phase $\phi$ and $\theta$, with (a) anti-symmetry detuning $\Delta_a = -\Delta_b = 0.5J$ and (b) symmetry detuning $\Delta_a = \Delta_b = 0.5J$. The other parameters used are $p=\Gamma=J$.}
    \label{purity}
\end{figure*}

\section{Generation of Pure States of qubits and Transparency} \label{transSec}
When the qubit spacing is commensurate with the drive wavelength and the detunings are antisymmetric, a waveguide-coupled array with an even number of qubits admits a pure steady state \cite{pichler2015quantum}. Perfect transparency in related waveguide-QED settings has likewise been identified \cite{mukhopadhyay2020transparency,miao2025transparency}. 
Here we extend these results to a two-qubit, nonchiral waveguide with DMI. 
(i) For symmetric detuning (\(\Delta_a=\Delta_b\)), a pure steady state emerges only in the presence of the DMI. 
(ii) For antisymmetric detuning (\(\Delta_a=-\Delta_b\)), we derive the exact pure steady state that persists with exchange. 
In both cases, these pure steady states yield reciprocal transparency (unit transmission and zero reflection for forward and backward drive).

Guided by numerics (Fig.~\ref{purity}), which reveal pure steady states for both symmetric and antisymmetric detunings, we now identify their structure analytically. To expose the jump structure, we rewrite the master equation as
\begin{equation}
\begin{aligned}
\dot\rho&=-i\!\left[H+\Gamma\sin\theta\!\left(S_1^+S_2^-+S_2^+S_1^-\right),\rho\right]\\
&+\Gamma \,\mathcal L_{c_\rightarrow}\rho \;+\; \Gamma \,\mathcal L_{c_\leftarrow}\rho,
\end{aligned}
\end{equation}
with the Lindblad superoperator
\(\mathcal L_\xi(X)\equiv \xi X\xi^\dagger-\tfrac12\{\xi^\dagger\xi,X\}\)
and the bidirectional jump operators
\begin{equation}
c_\leftarrow=S_a^-+e^{i\phi}S_b^-,
\qquad
c_\rightarrow=S_a^-+e^{-i\phi}S_b^-.
\end{equation}
By the cyclicity of the trace, the Hamiltonian term leaves the purity invariant. The dissipators yield
\begin{equation}
\frac{d}{d t} \operatorname{Tr} \rho^2=2 \Gamma\sum_\mu \left[\operatorname{Tr}\left(\rho c_\mu \rho c_\mu^{\dagger}\right)-\operatorname{Tr}\left(\rho^2 c_\mu^{\dagger} c_\mu\right)\right],
\end{equation}
with $\mu\in\{\rightarrow,\leftarrow\}$ labeling the right- and left-propagating waveguide channels. Evaluating the right-hand side on a pure state \(\rho=|P\rangle\!\langle P|\) gives
\(
-2\Gamma\sum_{\mu}\Big(\langle c_\mu^\dagger c_\mu\rangle
-|\langle c_\mu\rangle|^2\Big)\le 0.
\)
Thus a pure state can be stationary only if, for each channel \(\mu\), 
\(\langle c_\mu^\dagger c_\mu\rangle=|\langle c_\mu\rangle|^2\), which is equivalent to the eigenstate conditions,
\begin{equation}
c_\mu|P\rangle=\lambda_\mu|P\rangle\qquad(\mu=\leftarrow,\rightarrow).
\label{pure}
\end{equation}

In our two–qubit setting, the jump operators $c_\mu$ ($\mu=\leftarrow,\rightarrow$) are finite sums of spin–$\tfrac12$ lowering operators, hence nilpotent in the 4D space ($c_\mu^3=0$). A nilpotent operator has only the eigenvalue $0$, so a pure steady state $|P\rangle$ must satisfy the conditions
$c_\mu|P\rangle=0$ for both directions. 
For the general state
$|P\rangle=a|ee\rangle+b|eg\rangle+c|ge\rangle+d|gg\rangle$,
since $c_\mu|ee\rangle\propto|ge\rangle+e^{\pm i\phi}|eg\rangle\neq 0$, stationarity forces $a=0$.
Acting on the single–excitation amplitudes yields the pair of constraints $b+e^{i\phi}c=0$, $e^{-i\phi}b+c=0$.
These are compatible only if $e^{2i\phi}=1$, i.e. $\phi=n\pi$ ($n\in\mathbb Z$); otherwise $b=c=0$ and $|P\rangle=d|gg\rangle$ is the only dark state.
Thus, nontrivial pure steady states exist precisely at the phase-matched separations
\begin{equation}
\phi=n\pi, 
\end{equation}
with
\begin{equation}
|P\rangle=d\,|gg\rangle+\beta\,|D_n\rangle,
\qquad
|D_n\rangle=\frac{|eg\rangle-(-1)^n|ge\rangle}{\sqrt2},
\end{equation}
and the bright orthogonal partner $|B_n\rangle=\frac{|eg\rangle+(-1)^n|ge\rangle}{\sqrt2}$. This general result matches the numerical observation in Fig.~\ref{purity} that completely pure steady states occur only at $\phi=n\pi$.
\begin{figure*}[tbp]
    \includegraphics[width=0.48\textwidth]{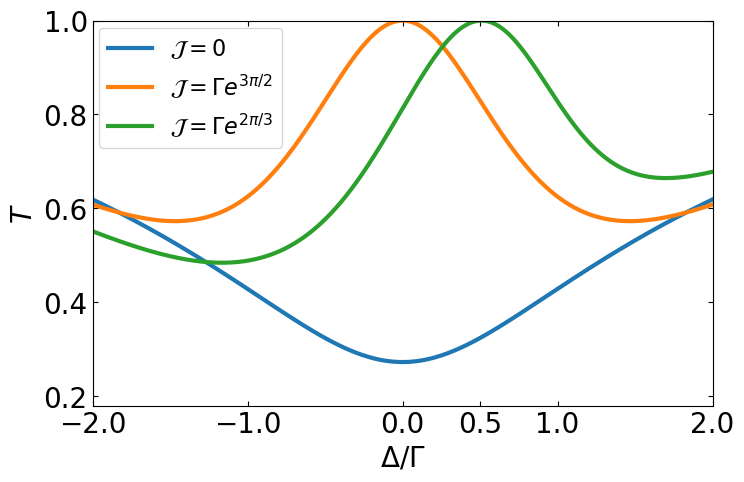}
    \caption{The transmission $T$ as a function of symmetry detuning $\Delta = \Delta_a = \Delta_b$ with different exchanges, $\mathcal{J}=0,J e^{3i\pi/2}, J e^{2i\pi/3}$. The other parameters used are $\phi=\pi$, $p=\Gamma=J$.}
    \label{trans}
\end{figure*}
\paragraph*{Symmetric detuning $\Delta_a=\Delta_b\equiv\Delta$ and nonzero DMI.}
A pure steady state also must be stationary under the coherent dynamics. That requires that the long-time state be an eigenstate of the Hamiltonian within this dark subspace $\mathrm{span}\{|gg\rangle,|D_n\rangle\}$; otherwise $H$ would rotate it into $|B_n\rangle$, destroying purity. We therefore seek
\begin{equation}
H|P\rangle=E\,|P\rangle,
\label{purestate}
\end{equation}
under single–port drive (forward) with amplitude $\alpha$.

In the $\{|gg\rangle,|D_n\rangle,|B_n\rangle\}$ basis the ingredients are: (i) the detuning term gives $-\Delta\,|gg\rangle$; (ii) the coherent drive couples $|gg\rangle\leftrightarrow|B_n\rangle$ with matrix element $\sqrt2\,k\,\alpha$ but does not couple to $|D_n\rangle$; and (iii) the complex exchange
\(
H_I=Je^{i\theta}S_1^+S_2^-+Je^{-i\theta}S_1^-S_2^+
\)
acts in the single–excitation manifold as
\begin{equation}
\begin{aligned}
&\langle D_n|H_I|D_n\rangle=-(-1)^n J\cos\theta,\\
&\langle B_n|H_I|D_n\rangle=-\,i(-1)^nJ\sin\theta .
\end{aligned}
\end{equation}
Acting on $|P\rangle$ and projecting on the three basis vectors yields
\begin{equation}
\begin{aligned}
&Ed=-\Delta d,\\ 
&E\,\beta=-(-1)^n J\cos\theta\,\beta,\\
&i\sqrt2k\alpha d- i\,(-1)^n J\sin\theta\,\beta=0.
\end{aligned}
\end{equation}
Nontrivial solutions with \(\beta\neq0\) require the phase–detuning lock
\begin{equation}
\Delta = (-1)^n J\cos\theta ,
\label{theta}
\end{equation}
which also sets \(E=-\Delta\), ensuring no energy mismatch between $|gg\rangle$ and $|D_n\rangle$ in the rotating frame. 
The third equation cancels the bright leakage, $\beta=\frac{\sqrt2\,k\,\alpha}{(-1)^nJ\sin\theta}d$, so $H$ leaves the state inside the dark manifold. 
Thus, for symmetric detuning and $\phi=n\pi$, the normalized long-time pure eigenstate is
\begin{equation}
|P\rangle=\frac{1}{\sqrt{2k^2\alpha^2+J^2-\Delta^2}}[\sqrt2k\alpha|D_n\rangle+(-1)^nJ\sin\theta|gg\rangle],
\label{purestate1}
\end{equation}
with $\theta$ determined by Eq.~(\ref{theta}). Figure~\ref{purity}(b) confirms the phase locking for the chosen detuning $\Delta=J/2$: for even $n$ the pure state appears at $\theta=\pi/3,\ 5\pi/3$, while for odd $n$ it shifts to $\theta=2\pi/3,\ 4\pi/3$. This matches the analytic condition Eq.~(\ref{theta}).

The DMI part $J\sin\theta$ provides a coherent coupling between the dark and bright single-excitation modes, ($|D_n\rangle\!\leftrightarrow\!|B_n\rangle$ coupling), while the drive injects amplitude from $|gg\rangle$ into $|B_n\rangle$ with strength $g$. By choosing $\theta$ (and the detuning as above), we can make the $|B_n\rangle$ component vanish exactly. In contrast, without DMI this cancellation is impossible: drive populates $|B_n\rangle$, leakage out of the dark manifold occurs, and the steady state is mixed.

\paragraph*{Antisymmetric detuning \(\Delta_a=-\Delta_b\equiv\Delta\) and nonzero DMI.}
Similarly, for the antisymmetric–detuning setting under single–port drive of amplitude $\alpha$, we seek the pure state of the form in Eq.~(\ref{purestate}) with $|D_n\rangle$ and $|B_n\rangle$ defined at $\phi=n\pi$. The antisymmetric detuning term $ \Delta(S_a^z - S_b^z) $ has no diagonal matrix element on $|gg\rangle$ and, within the single–excitation manifold, it contributes to a coherent coupling between $|D_n\rangle$ and $|B_n\rangle$:
$\langle B_n|\Delta(S_a^z-S_b^z)|D_n\rangle=\Delta$.

Projecting \(H|P\rangle\) onto \(|gg\rangle,|D_n\rangle,|B_n\rangle\) yields the three conditions
\begin{equation}
\begin{aligned}
&Ed=0,\\
&E\,\beta=-(-1)^n J\cos\theta\,\beta,\\
&i \sqrt2k\alpha d+ \big(\Delta - i\,(-1)^n J\sin\theta\big)\beta=0.
\end{aligned}
\end{equation}

\begin{figure*}[htbp]
    \includegraphics[width=0.96\textwidth]{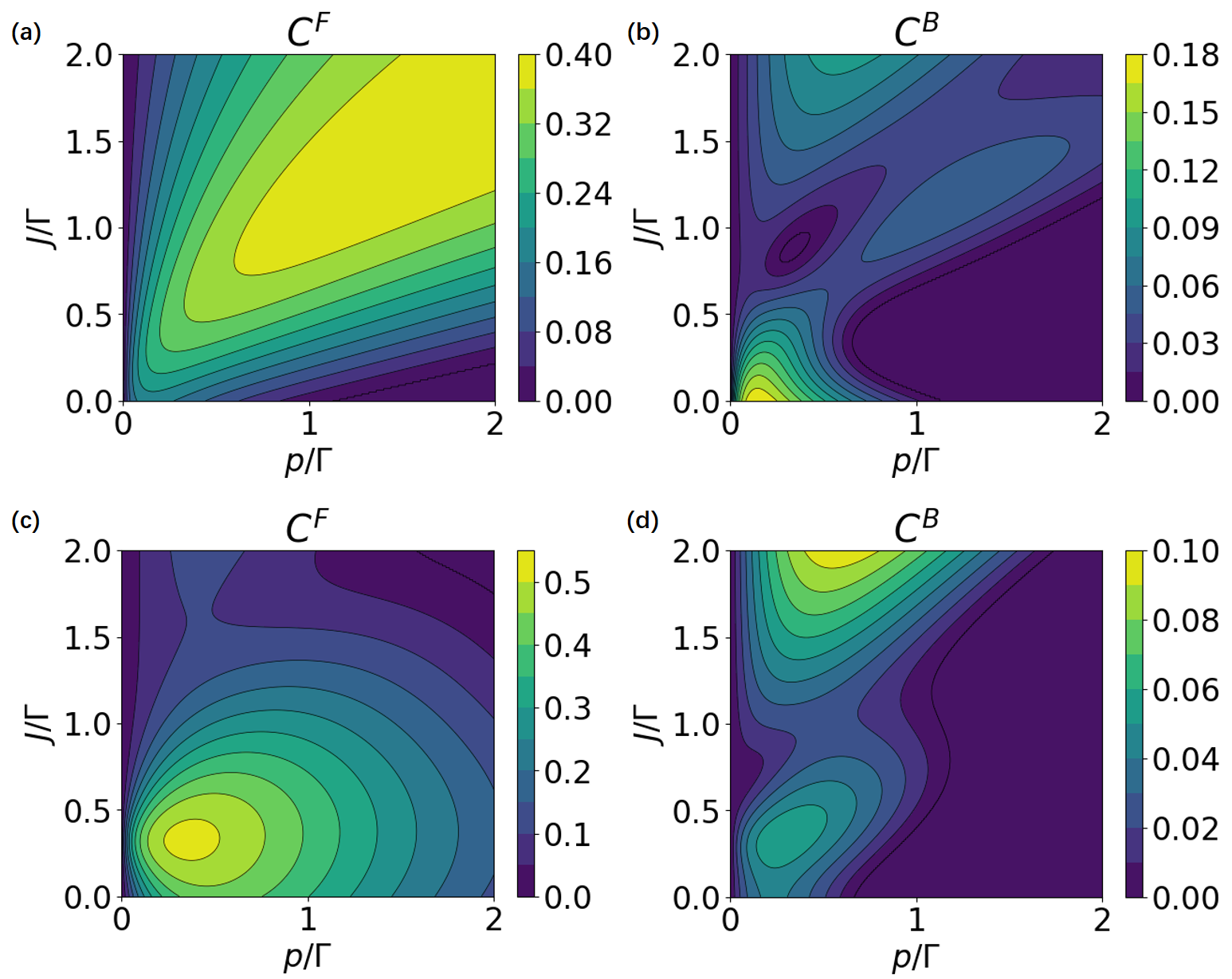}
    \caption{Contour plots of forward and backward of steady-state quantum entanglement, quantified by concurrence (a,c) $C^F$ and (b,d) $C^B$, as functions of the exchange strength $J$ and driving strength $p$, with (a,b) symmetric detuning and (c,d) antisymmetric detuning. The parameters are (a,b) $\phi=9\pi/50$, $\theta=9\pi/25$, $\Delta_a = \Delta_b = 0.5\Gamma$; (c,d) $\phi=\pi/10$, $\theta=9\pi/10$, $\Delta_a = -\Delta_b = 0.5\Gamma$. These phases are chosen to produce pronounced nonreciprocity.}
    \label{entangle}
\end{figure*}

For nontrivial solutions, the first two imply
\begin{equation}
\theta=\frac{\pi}{2},\;\frac{3\pi}{2},
\label{theta2}
\end{equation}
with nonzero $J$. The third equation fixes the superposition weight, $\beta=\frac{\,\sqrt2\,k\,\alpha}{(-1)^n \,J\sin\theta+i\Delta}d$. Thus, for antisymmetric detuning and \(\phi=n\pi\) the long–time state
\begin{equation}
\begin{aligned}
|P\rangle &= \frac{1}{\sqrt{2k^2\alpha^2+J^2+\Delta^2}}\\&\times[\sqrt2k\alpha|D_n\rangle+((-1)^nJ\sin\theta+i\Delta)|gg\rangle]
\end{aligned}
\label{purestate2}
\end{equation}
is pure. Unlike the symmetric–detuning case, here the detuning \(\Delta\) is not locked to \(J\); instead the DMI phase is pinned to \(\theta=\pi/2,3\pi/2\) as shown in Fig. \ref{purity}(a).

\paragraph*{Purity leads to transparency.}
Using the input–output relations, it is convenient to identify the jump operators. Eqs. (\ref{inoutput}) may then be written as
\begin{equation}
\begin{aligned}
&\varepsilon_{1\leftarrow}=e^{i\phi}\varepsilon_{2\leftarrow}-kc_{\leftarrow},\\
&\varepsilon_{2\rightarrow}=e^{i\phi}\varepsilon_{1\rightarrow}-ke^{i\phi}c_{\rightarrow}.
\end{aligned}
\end{equation}

For the pure steady states derived above one has
\(
c_{\leftarrow}|P\rangle=0=c_{\rightarrow}|P\rangle
\),
hence under forward single–port drive
\((\varepsilon_{1\rightarrow}=\alpha,\ \varepsilon_{2\leftarrow}=0)\), $\langle \varepsilon_{1\leftarrow}\rangle=0$, 
$\langle \varepsilon_{2\rightarrow}\rangle=e^{i\phi}\alpha$,
and
\(
\langle \varepsilon_{1\leftarrow}^\dagger\varepsilon_{1\leftarrow}\rangle=0,\;
\langle \varepsilon_{2\rightarrow}^\dagger\varepsilon_{2\rightarrow}\rangle=|\alpha|^{2}.
\)
Therefore the coherent amplitude transmission and the total transmission are
\begin{equation}
|t|=\Big|\frac{\langle \varepsilon_{2\rightarrow}\rangle}{\alpha}\Big|=1,
\qquad
T=\frac{\langle \varepsilon_{2\rightarrow}^\dagger\varepsilon_{2\rightarrow}\rangle}{|\alpha|^{2}}=1,
\end{equation}
and exchanging the ports, the same holds for backward drive. 
This reciprocity can be viewed as a symmetry: interchanging the drive and detection ports is equivalent to sending \(\theta\to 2\pi-\theta\), which leaves the analytic conditions Eq. (\ref{theta}) and Eq. (\ref{theta2}) invariant.

Fig.~\ref{trans} displays the steady–state transmission for symmetric detuning
\(\Delta_a=\Delta_b=\Delta\), comparing a reference (\(\mathcal J=0\), blue) with two exchange settings \(\mathcal J=\Gamma e^{3i\pi/2}\) (orange) and \(\mathcal J=\Gamma e^{2i\pi/3}\) (green).  
When \(\phi=n\pi\), \(\mathcal J=0\) and the detunings are symmetric, the system has a qubit-exchange (permutation) symmetry that decouples the single-excitation dark/bright manifolds \(|D_n\rangle\) and \(|B_n\rangle\). In particular, with \(\phi=\pi\) (odd \(n\)) one has \(|D_n\rangle=|s\rangle\) and \(|B_n\rangle=|a\rangle\), where $|s\rangle$ is the symmetric triplet and $|a\rangle$ the antisymmetric singlet. The Hamiltonian contains no term that mixes $|s\rangle$ and $|a\rangle$; the drive couples $|gg\rangle\leftrightarrow|a\rangle$ while $|s\rangle$ remains dynamically isolated. The equations of motion then give \(\dot\rho_{ss}=0\), so if the system starts in \(|gg\rangle\) one has \(\rho_{ss}(t)\equiv 0\) for all \(t\). Consequently, there is no transparency, and the transmission exhibits a minimum at \(\Delta=0\). An analytic expression for the transmission in this \(\mathcal J=0\) reference is obtained:

\begin{equation}
T=
\frac{\alpha ^2 \left(\Delta ^4+k^4 \left(3 \alpha ^4+\Delta ^2\right)+4 \alpha ^2 \Delta ^2 k^2\right)}{\Delta ^4+4 k^8+3 \alpha ^4 k^4+5 \Delta ^2 k^4+4 \alpha ^2 k^2 \left(\Delta ^2+k^4\right)} .
\end{equation}

Switching on DMI coherently reintroduces the \(|D_n\rangle\!\leftrightarrow\!|B_n\rangle\) coupling.  
For \(\mathcal J=\Gamma e^{3i\pi/2}\) the central dip is converted into unit transmission at \(\Delta=0\), consistent with the symmetric-detuning pure-state condition in Eq. (\ref{theta}).  
For \(\mathcal J=\Gamma e^{2i\pi/3}\), Eq. (\ref{theta}) predicts a transparency peak at \(\Delta=J/2\); with \(J=\Gamma\) this occurs at \(\Delta/\Gamma=1/2\), exactly as seen in Fig.~\ref{trans}.  
These phase–programmable peaks confirm that DMI supplies the missing dark–bright coupling under symmetric detuning and enables power–independent, unit transparency at the predicted \((\Delta,\theta)\).

\section{Nonreciprocal Quantum Entanglement produced by DMI} \label{entSec}
\begin{figure*}[tbp]
    \includegraphics[width=0.96\textwidth]{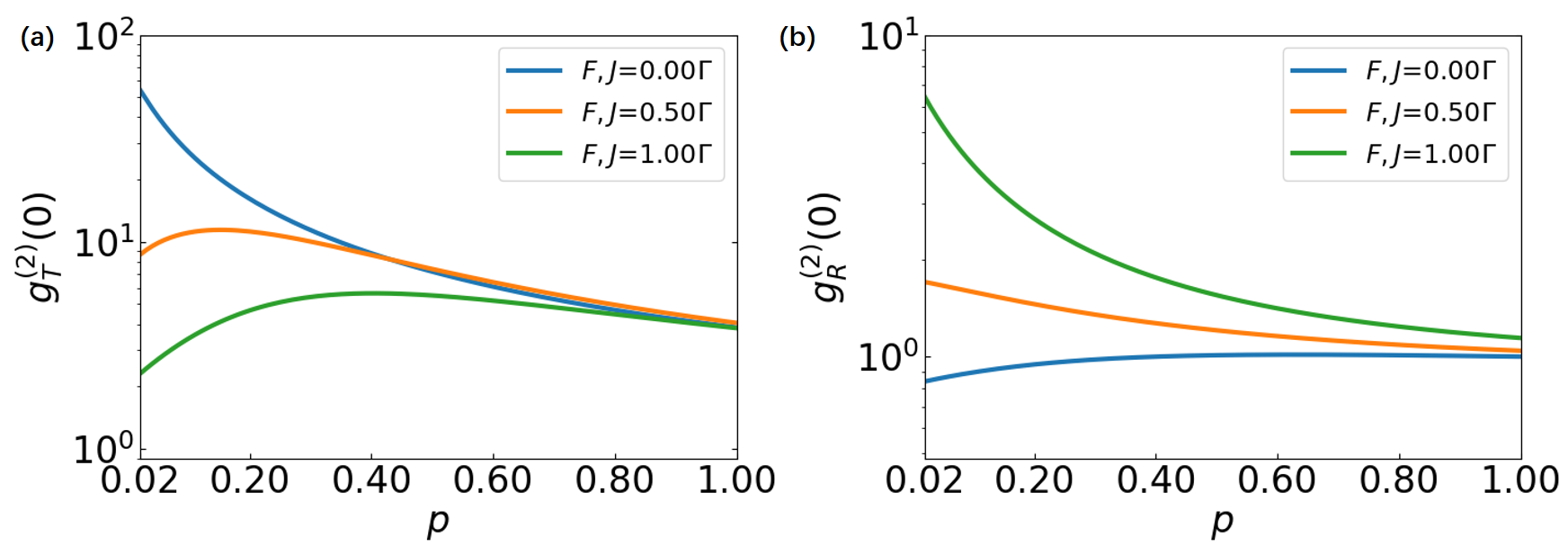}
    \caption{Normalized second-order intensity correlations $g^{(2)}(0)$ of the (a) transmitted and (b) reflected fields versus drive strength $p$ at $\phi=3\pi/4$ for $J/\Gamma=0,\ 0.5,\ 1$. The curves are plotted from $p=0.02\Gamma$. Parameters: $\theta=5\pi/4$, $\Delta_a=-\Delta_b=0.5\Gamma$. All is driven in forward propagation.}
    \label{superb}
\end{figure*}
Typically, nonreciprocity is discussed in terms of transport—unequal forward/backward transmission under otherwise identical conditions. In principle, the same interference mechanism that skews transport can also skew genuinely quantum features. Here we demonstrate nonreciprocity in quantum entanglement: with all parameters fixed and only the driven port exchanged, the steady‐state two–qubit entanglement differs. 

We quantify bipartite entanglement by the Wootters concurrence \cite{wootters1998entanglement},
\begin{equation}
C(\rho)=\max\{0,\lambda_1-\lambda_2-\lambda_3-\lambda_4\},
\end{equation}
where \(\{\lambda_i\}\) are the square roots of the eigenvalues of \(\rho\,\tilde\rho\) arranged in nonincreasing order. The “spin–flipped” matrix \(\tilde\rho\) is
\(
\tilde\rho=(\sigma_y\!\otimes\!\sigma_y)\,\rho^{*}\,(\sigma_y\!\otimes\!\sigma_y),
\)
with \(\sigma_y\) the Pauli \(y\) matrix and \(\rho^{*}\) the complex conjugate of \(\rho\) taken in the computational basis 
\(\{|ee\rangle,|eg\rangle,|ge\rangle,|gg\rangle\}\).
By construction \(0\le C(\rho)\le 1\). We denote by \(C^{F}\) (\(C^{B}\)) the concurrence of the steady state under forward (backward) single-port driving.

A clear directional asymmetry emerges. To make the effect pronounced, we choose phases \(\phi=9\pi/50\), \(\theta=9\pi/25\) for symmetric detuning; \(\phi=\pi/10\), \(\theta=9\pi/10\) for antisymmetric detuning. For symmetric detuning, forward drive [Fig.~\ref{entangle}(a)] exhibits a broad high-\(C\) region with a maximum \(C^{F}\!\approx\!0.40\) near \((p/\Gamma,J/\Gamma)\!\approx\!(1.71,2)\),
while the backward case [Fig.~\ref{entangle}(b)] is much weaker (peak \(C^{B}\!\approx\!0.18\)) and shows extended low-\(C\) domains at moderate power. For antisymmetric detuning the contrast is even stronger: forward drive [Fig.~\ref{entangle}(c)] yields a wide high-\(C\) lobe with \(C^{F}\!\approx\!0.52\) near \((0.33,0.34)\),
whereas backward drive [Fig.~\ref{entangle}(d)] reaches only \(C^{B}\!\approx\!0.10\). Thus, over most of the \((p,J)\) plane we find \(C^{F}\!\neq\!C^{B}\), i.e., DMI-induced nonreciprocity in steady-state entanglement. Note that recent superconducting proposals discuss nonreciprocal entanglement in transient domain by preparing one qubit initially excited, driving only a single qubit, and working in the cascaded–systems limit with a single collective jump operator \cite{ren2025nonreciprocal}. 
\begin{figure*}[tbp]
    \includegraphics[width=0.96\textwidth]{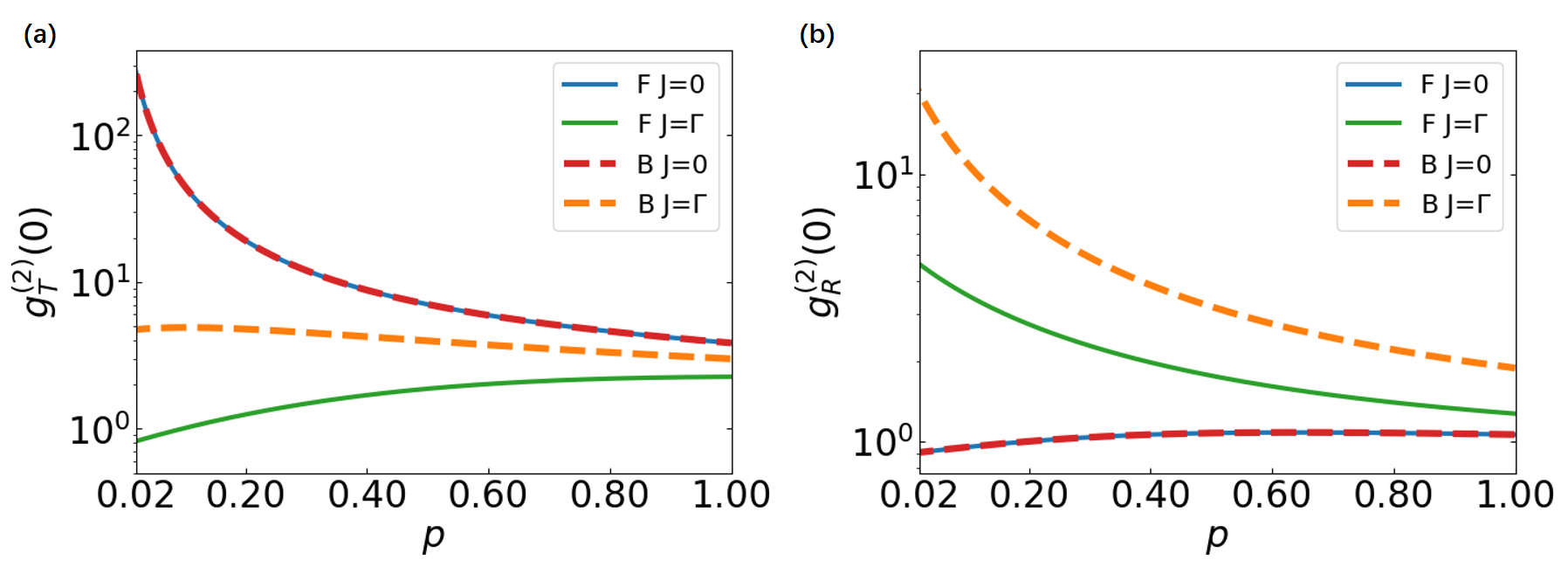}
    \caption{Normalized second-order intensity correlations $g^{(2)}(0)$ for forward (F) and backward (B) driving with exchange $\mathcal J=0,\Gamma e^{5\pi/4}$. Curves are plotted from $p=0.02\Gamma$. Parameters: $\phi=3\pi/4$, $\Delta_a=\Delta_b=0.5\Gamma$.}
    \label{superb2}
\end{figure*}

Finally, we examine how the exchange magnitude $J$ controls entanglement. 
At the pure–state transparency points the concurrence is analytic: \begin{equation}
C(|P\rangle)=
\frac{2k^{2}\alpha^{2}}{\,2k^{2}\alpha^{2}+J^{2}\mp \Delta^{2}\,}\,,
\end{equation}
with upper (minus) sign for symmetric detuning and lower (plus) sign for antisymmetric detuning. For symmetric detuning, Eq. (\ref{theta}) admits a solution only if $J\ge|\Delta|$. When $J<|\Delta|$ no solution exists and no pure steady state is possible. Approaching the boundary $J\rightarrow|\Delta|$ forces $\cos\theta\rightarrow\pm1$ (hence $\sin\theta\rightarrow0$), so the pure state collapses onto the single-excitation dark state (the $|gg\rangle$ component vanishes) and $C(|P\rangle)\rightarrow1$. However, the relaxation into this state becomes arbitrarily slow: the coherent coupling that feeds the dark manifold scales as $J\sin\theta$, so the approach time diverges as $J\sin\theta\rightarrow0$.

In both detuning patterns, for fixed power, increasing $J$ reduces the pure-state concurrence. 
Away from the pure–state conditions, however, $C$ becomes nonmonotonic in $J$: as seen in Fig.~\ref{entangle}, it can increase with $J$—indicating that the exchange opens an effective entangling channel. This $J$-dependent rise–and–fall, with different magnitudes for forward and backward drive, offers a direct and tunable handle for programming steady–state entanglement and its nonreciprocity. Experimentally, entanglement under forward and backward driving can be quantified separately, most directly via two-qubit state tomography of the steady state \cite{steffen2006measurement,dicarlo2009demonstration}.

\section{Nonreciprocal photon correlations by DMI} \label{SuperbunchSec}

Photon antibunching (\(g^{(2)}(0)<1\))—the suppression of coincident detections at zero delay—is a hallmark of single-emitter quantum optics. A modern view attributes antibunching to quantum interference between two-photon scattering pathways, coherent and incoherent. If the coherent scattered component is spectrally filtered out, the interference channel is removed and the remaining fluorescence exhibits strong bunching \cite{masters2023simultaneous}. One-dimensional waveguides are ideally suited to resolve these components: their well-defined spatial modes provide direct access to both field amplitudes and powers in transmission and reflection, enabling a clean probe of two-photon statistics. In this section, we exploit this platform to investigate how DMI reshapes photon correlations.

In a waveguide-QED system with weakly coupled emitters, the interaction with the propagating continuum can generate strongly correlated photon streams \cite{mahmoodian2018strongly}. A striking manifestation is superbunching, where the zero-delay second-order correlation exceeds the thermal value, \(g^{(2)}(0)> 2\), implying that photons prefer to appear in pairs rather than independently. We quantify the statistics in a given detection channel through
\begin{equation}
g^{(2)}(0)=
\frac{\big\langle \varepsilon_i^\dagger\,\varepsilon_i^\dagger\,
\varepsilon_i\,\varepsilon_i \big\rangle}
{\big|\langle \varepsilon_i^\dagger \varepsilon_i\rangle\big|^2},
\end{equation}
where \(\varepsilon_i\) is the output-field operator obtained from the input–output relations, and $i\in\{1\rightarrow,2\leftarrow\}$ labels the output channels. Below we show that DMI provides a tunable route to superbunching in a nonchiral waveguide. 

Fig.~\ref{superb2} plots the zero–delay intensity correlations in transmission, \(g^{(2)}_{T}(0)\), and reflection, \(g^{(2)}_{R}(0)\), versus the drive power \(p\) (\(\phi=3\pi/4\) and \(\Delta_a=\Delta_b=0.5\,\Gamma\)). Solid (dashed) curves correspond to forward (backward) single–port driving; we compare a reciprocal reference \((\mathcal J=0)\) with a complex exchange \((\mathcal J=\Gamma e^{5\pi i/4})\). For \(\mathcal J=0\) the forward and backward traces coincide in both panels, as expected for a nonchiral waveguide under symmetric detuning. With DMI, a clear directional pattern emerges and the port response splits:
(i) in transmission [Fig.~\ref{superb2}(a)], both forward and backward \(g^{(2)}_{T}(0)\) lie below the reciprocal reference across the full range of \(p\); 
(ii) in reflection [Fig.~\ref{superb2}(b)], both forward and backward \(g^{(2)}_{R}(0)\) are enhanced above the reference, with the backward drive yielding the much larger correlation—superbunching that remains largest even at strong driving.

Thus, DMI redistributes two-photon correlations: it suppresses them in transmission while amplifying them in reflection, and it does so asymmetrically with respect to drive direction, with the backward port exhibiting the stronger correlation (superbunching).

Fig.~\ref{superb} reveals a complementary J–dependence of photon statistics under forward drive (\(\phi=3\pi/4\), \(\theta=5\pi/4\), \(\Delta_a=-\Delta_b=0.5\Gamma\)).  
In transmission [Fig.~\ref{superb}(a)], the zero-delay correlations \(g^{(2)}_{T}(0)\) decrease as the exchange strength is raised from \(J=0\) to \(0.5\Gamma\) and \(\Gamma\); i.e., stronger DMI suppresses pair correlations in the transmitted field.  
In reflection [Fig.~\ref{superb}(b)], the trend inverts: \(g^{(2)}_{R}(0)\) increases with \(J\), indicating that the DMI shifts two-photon correlations from transmission to reflection. Similar results are obtained for backward driving. Taken together with Fig.~\ref{superb2}, these data show that the complex exchange redistributes bunching between the two ports in a controllable way.

\section{CONCLUSION}
\label{conclusion}
We have analyzed two qubits in a bidirectional, nonchiral waveguide coupled by a complex exchange whose antisymmetric (DMI) component carries a tunable phase. Solving the full master equation with input–output theory, we showed that the DMI component alone generates sizable forward–backward splittings in both the coherent and incoherent parts of the transmission, programmable by the drive power and by the phases \((\phi,\theta)\).

Antisymmetric exchange can be engineered within waveguide QED by tailoring how emitters couple to multiple waveguide channels. In the fully chiral (cascaded) limit, the waveguide-mediated coherent coupling is purely directional and antisymmetric, while in a bidirectional (nonchiral) waveguide the induced coherent exchange is purely symmetric. Combining channels therefore provides independent control over the symmetric and antisymmetric contributions to the effective complex exchange $\mathcal J$. Since both $\mathcal J$ and the collective decay terms originate from the same underlying waveguide-mediated photon exchange processes, they can be jointly engineered by the channel couplings—for example by adjusting the emitter positions relative to the waveguide.

At phase-matched spacings $\phi = n\pi$, the dynamics isolate a single dark mode: any pure steady state must lie in its span and therefore yields reciprocal unit transparency. For symmetric detuning, such a state exists only when the DMI supplies the coherent dark–bright mixing; for antisymmetric detuning, a pure state persists with $\theta = \pi/2,\ 3\pi/2$ while $\Delta$ remains free. In both cases the state is dark to both jump channels, converting the no-DMI transmission dip into phase-programmable unity peaks.

Beyond transport, the same phase bias imprints directional entanglement: over broad regions of ($p$, $J$) we find $C^{F}\!\neq\!C^{B}$. At transparency points the steady state is pure and $C$ decreases with increasing $J$ at fixed power; away from those points the concurrence becomes nonmonotonic in $J$, evidencing a DMI-enabled entangling pathway. Finally, the DMI redistributes two-photon correlations between ports and directions, yielding tunable superbunching $g^{(2)}(0)\!>\!2$ with a pronounced backward–forward asymmetry.

Altogether, a complex (DMI) exchange provides a compact, phase-programmable knob to engineer nonreciprocity, transparency, entanglement, and photon statistics in reciprocal waveguides. Our findings open promising directions for the realization of quantum photonic devices such as isolators, routers, and sources of superbunching light—without relying on chiral waveguides.

\section{Acknowledgments}
We are grateful for the support of the Robert A Welch Foundation (Grant No. A-1943-20240404) and the Department of Energy, Fusion Energy Science, Award No. DE-SC0024882, IFE-STAR. Z. Z. is supported by the Heep Graduate Fellowship.

\appendix
\section{Time reversal and exchange interaction}
\label{timerev}
In the main text, we consider two qubits coupled by a complex exchange interaction. The system Hamiltonian
$H(\theta;\varepsilon_{1\to},\varepsilon_{2\leftarrow})$ [Eq.~(\ref{Hamiltonian})] is written in the rotating
frame of the coherent drive, where $\varepsilon_{1\to}$ ($\varepsilon_{2\leftarrow}$) denotes the complex
amplitude of the incident field injected from the left (right) port.

For a two-level system (effective pseudo-spin-$1/2$) we represent time reversal by an antiunitary operator
$\mathcal T$ that includes complex conjugation, so that $\mathcal T i \mathcal T^{-1}=-i$.
In the $\{|g\rangle,|e\rangle\}$ basis, $\mathcal T$ acts as complex conjugation on scalar coefficients
(and leaves the spin operators invariant up to an irrelevant phase). Applying $\mathcal T$
to the driven Hamiltonian yields
\begin{equation}
\mathcal T\,H(\theta;\varepsilon_{1\to},\varepsilon_{2\leftarrow})\,\mathcal T^{-1}
= H(-\theta;\varepsilon_{2\leftarrow}^*,\varepsilon_{1\to}^*),
\label{eq:TR_H}
\end{equation}
which expresses two standard consequences of time reversal: (i) the complex exchange phase changes sign,
$\theta\to-\theta$; and (ii) externally prepared coherent source fields are complex conjugated and their
propagation directions are reversed, i.e., $\varepsilon_{1\to}\mapsto \varepsilon_{2\leftarrow}^*$ and
$\varepsilon_{2\leftarrow}\mapsto \varepsilon_{1\to}^*$.
Accordingly, the exchange part of the Hamiltonian is time-reversal invariant only for $\theta=0$ or $\pi$,
where the exchange amplitudes are purely real (equivalently, the DMI component $J\sin\theta$ vanishes).
For $\sin\theta\neq 0$, fixing a nonzero $\theta$ breaks time-reversal symmetry.

\section{Unitary qubit/port exchange as an alternative viewpoint}
\label{qubitex}
One may also view Eqs.~(\ref{Hamiltonian}-\ref{inoutput}) from a unitary
relabeling symmetry when the two qubits are identical.
Define the qubit-permutation operator $\mathcal P$ by
\begin{equation}
\mathcal P S_a^\mu \mathcal P^{-1}=S_b^\mu,\qquad
\mathcal P S_b^\mu \mathcal P^{-1}=S_a^\mu,
\qquad \mu\in\{z,+,-\}.
\end{equation}
If $\Delta_a=\Delta_b$ and the qubits couple identically to the waveguide and loss channels, then swapping
$a\leftrightarrow b$ interchanges the two exchange terms and hence flips $\theta\to-\theta$.
Swapping the qubits must be accompanied by exchanging the drive ports, $\mathcal P \varepsilon_{1\to} \mathcal P^{-1}=\varepsilon_{2\leftarrow}$, which yields
\begin{equation}
\mathcal P\,H(\theta;\varepsilon_{1\to},\varepsilon_{2\leftarrow})\,\mathcal P^{-1}
=H(-\theta;\varepsilon_{2\leftarrow},\varepsilon_{1\to}).
\label{eq:P_swap_H}
\end{equation}
Under the same identical-qubit conditions, the dissipator in Eq.~(\ref{master}) is invariant under
$a\leftrightarrow b$.
Hence, if $\rho^{F}_{\rm ss}(\theta)$ is the steady state under forward drive
$(\varepsilon_{1\to},\varepsilon_{2\leftarrow})=(\alpha,0)$, then
\begin{equation}
\rho^{B}_{\rm ss}(-\theta)=\mathcal P\,\rho^{F}_{\rm ss}(\theta)\,\mathcal P^{-1}
\end{equation}
is the steady state under backward drive $(0,\alpha)$.
Finally, the input--output relations are preserved under the same combined qubit/port exchange, which yields
\begin{equation} 
\langle \varepsilon_{2\rightarrow}^\dagger \varepsilon_{2\rightarrow}\rangle^F_{\theta}=\langle \varepsilon_{1\leftarrow}^\dagger \varepsilon_{1\leftarrow}\rangle^B_{-\theta},\qquad  \langle\varepsilon_{2\rightarrow}\rangle^F_{\theta}=\langle  \varepsilon_{1\leftarrow}\rangle^B_{-\theta}.
\end{equation} 

When $\mathcal J=0$ and the propagation phase satisfies $\phi=n\pi$, the qubit-exchange (permutation) symmetry occurs under single-sided driving, which is discussed in
Sec.~\ref{transSec}.
Finally, if the two qubits are detuned ($\Delta_a\neq\Delta_b$), nonreciprocity
can also be induced by quantum nonlinearity even without a DMI component, as discussed in Ref.~\cite{rosario2018nonreciprocity}.

\section{Coherent transmission in the weak-drive limit}
\label{weaklimit}
In the weak-drive limit ($p\to 0$), the steady state is well described within the linear response sector: saturation is negligible, the incoherent contribution vanishes, and transmission is governed by interference between the incident field and the coherently scattered field. As a result, the coherent transmission $T_c$ can vary sharply with system parameters because it is governed by interference, and it may even approach a near-zero minimum when the re-radiated field nearly cancels the transmitted incident field as seen from Fig. \ref{nonre}a. One may linearize the equations of motion [Eq.~(\ref{eq:cmt_dyn})] for the dipole coherences by taking the spins close to their ground-state values,
$\langle S_a^z\rangle=\langle S_b^z\rangle=-1/2$.
Solving the resulting steady-state linear system yields an analytic expression for the coherent transmission, which is reciprocal in linear response,
\begin{equation}
    \begin{aligned}
    T_c&=T_{c}^F=T_{c}^B \\
    &\!\!\!\!\!\!\!\!\!= \frac{|-J^2-2J\Gamma e^{-i\theta}\sin\phi+\Delta_a\Delta_b|^2}{|-J^2+2 i \Gamma  J e^{i \phi } \cos \theta +\Gamma ^2 e^{2 i \phi }-(\Gamma +i \Delta_a )(\Gamma +i \Delta_b)|^2}.
    \end{aligned}
    \label{weak_drive}
\end{equation}
For $J=0$, Eq.~(\ref{weak_drive}) reduces to a purely interference-controlled expression determined by
$(\Delta_a,\Delta_b,\phi)$, and for certain choices, the coherently scattered field nearly cancels the directly
transmitted input, yielding a very small $T_c$.
When $J\neq 0$, the additional terms $\propto J$ and $\propto J^2$ modify both the amplitude and phase of the
re-radiated field. $\theta$ enters Eq.~(\ref{weak_drive}) through both $e^{-i\theta}$ and $\cos\theta$, i.e., it tunes the
relative phase between detuning-mediated and exchange-mediated scattering pathways. Therefore, at fixed $J\neq 0$
the weak-drive coherent transmission can vary strongly as a function of $\theta$. In particular, for the parameters of Fig.~\ref{nonre} ($\Delta_a=\Delta_b=0.5\,\Gamma$, $\phi=9\pi/25$, $\theta=18\pi/25$), we obtain $T_{c}(J=\Gamma)=0.982$ while $T_{c}(J=0)=0.03$. For $\Delta_a=\Delta_b=0.5\,\Gamma$, $\phi=\pi/4$, $\theta=3\pi/4$, we find $T_{c}(J=\Gamma)=0.671$ and $T_{c}(J=0)=0.109$. Thus, at a weak drive, 
$\phi$ and $\theta$ can strongly modulate coherent transmission by tuning the relative phases of the competing scattering pathways.

\bibliography{refs}

\end{document}